\documentclass{epl}
\usepackage{amssymb}
\usepackage{amsmath}
\usepackage{citesort}
\usepackage{graphicx}
\title{Random maps in physical systems}
\shorttitle{Random maps in physical systems}
\shortauthor{.}
\author{L. Trujillo\inst{1}\inst{2}\inst{3}\footnote{E-mail: leo@pmmh.espci.fr}, J. J. Su\'arez\inst{3}, and J. A. Gonz\'alez\inst{2}\inst{3}}
\institute{ \inst{1} 
PMMH (CNRS UMR 7636), ESPCI, 10 rue Vauquelin 75231 Paris Cedex 05
France\\
\inst{2} 
  International Centre for Theoretical Physics (ICTP),
Trieste, Italy
\\
\inst{3} Centro de F\'{\i}sica,
IVIC, A.P. 21827, Caracas 1020-A, Venezuela}
\pacs{05.45.-a}{Nonlinear dynamics and nonlinear dynamical
systems}
\pacs{42.65.Sf}{Dynamics of nonlinear optical systems; optical
instabilities, optical chaos and complexity, and optical
spatio-temporal dynamics}
\pacs{05.45.Vx}{Communication using chaos}
\begin{document}

\maketitle

\begin{abstract}
We show that functions of type $X_n = P[Z^n]$, where $P[t]$ is a
periodic function and $Z$ is a generic real number, can produce
sequences such that any string of values $X_{s}, X_{s+1},
...,X_{s+m}$ is deterministically independent of past and future
values. There are no correlations between any values of the
sequence. We show that this kind of dynamics can be generated
using a recently constructed optical device composed of several
Mach--Zehnder interferometers. Quasiperiodic signals can be
transformed into random dynamics using nonlinear circuits. We
present the results of real experiments with nonlinear circuits
that simulate exponential and sine functions.
\end{abstract}

Recent experiments with electronic circuits have shown the
possibility of communication with chaos
\cite{PC90,PC91,DP93,CO93}.
The interesting question of communication with chaotic lasers has
also been discussed in \cite{VWR98}.
The fast dynamics displayed by optical systems offers the
possibility of communication at bandwidths of hundreds of
megahertz or higher. Very recently, Umeno et al. \cite{UAK01} have
proposed an optical device implementation of chaotic maps. They
rightly claim that the development of secure fiber--optic
communication systems can have a large impact on future
telecommunications. One problem in this area is constructing
all--optical devices for the transmission of high--bit--rate
signals with the appropriate security. Umeno et al. introduce
multi Mach--Zehnder (MZ) interferometers which implement a very
nice class of chaotic maps.
Other papers have shown that even chaotic communication systems
can be cracked if the chaos is predictable \cite{Ch96,Cer}.

In the present Letter we will show that using the same
experimental setup of Ref. \cite{UAK01} with some small
modifications and also other physical systems, it is possible to
construct random maps that generate completely unpredictable
dynamics.

S. Ulam and J. von Neumann \cite{UvN47,SU64} proved that the
logistic map $X_{n+1}=4X_n(1-X_n)$ can be solved using the
explicit function $X_n=\sin^2[\theta \pi 2^n]$. Other chaotic maps
are solvable exactly using, e.g., the functions
$X_n=\sin^2[\theta\pi k^n]$, $X_n=\cos[\theta\pi k^n]$, and other
functions of type $X_n=P[k^n]$, where $k$ is an integer
\cite{AR64,GF84,KF,Um97}. For instance,
$X_n=\sin^2[\theta \pi 3^n]$ is the exact general solution to the
cubic map $X_{n+1}=X_n(3-4X_n)^2$. The first--return maps for the
dynamics of these systems can be observed in Fig. 1.

\begin{figure}
\onefigure[width=8cm,height=4cm]{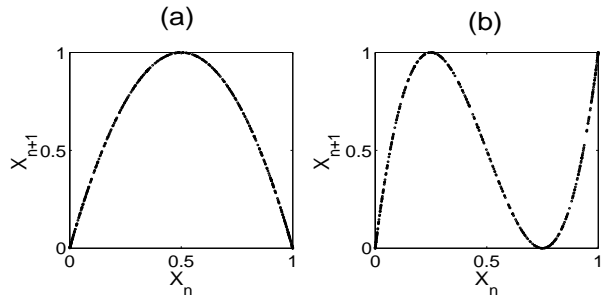}
\caption{First--return maps produced by functions
$X_n=\sin^2[\theta \pi 2^n]$ (a) and $X_n=\sin^2[\theta \pi 3^n]$
(b).}
\label{f.1}
\end{figure}

\begin{figure}
\onefigure[width=5cm,height=3cm]{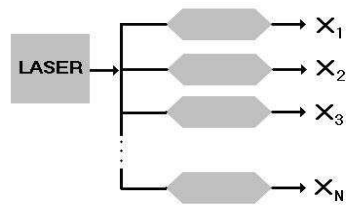}
\caption{Scheme of an experiment described in Ref.\cite{UAK01}
with $N$ Mach--Zehnder interferometers. In this experiment a
chaotic sequence is generated.}
\label{s.1}
\end{figure}

In Ref. \cite{UAK01} an optical circuit composed of $N$
MZ interferometers is presented. The scheme of this circuit is
shown in Fig. 2.
In this experimental setup, the input signal is divided into $N$
with equal power after passing an $1\times N$ coupler. The
intensity of light is measured at the output of the $n$--th
MZ interferometer using a power meter. Each intensity is defined
as $X_n$. The path length difference is given by $\Delta L (n)$.
The values of $\Delta L(n)$ satisfy the relationship
\begin{equation}
\Delta L(n+1) = m\Delta L(n),
\label{EQ1}
\end{equation}
where $m$ is an integer, $m>1$.

The transfer function at the output part of the $n$--th MZ
interferometer is given by the equation
%
$X_n =\sin^2\left[\pi \Delta L(n) r/\lambda\right],
$
%
%
where $\lambda$ is the wavelength of the light source and $r$ is
the effective refractive index of the optical paths of the MZ
interferometers. Thus the output powers $X_1,X_2,\ldots,X_N$
satisfy the equation
%
$X_n = \sin^2\left[ \pi r \Delta L(1) m^n/(\lambda m)\right]$.
%
%

The authors of Ref. \cite{UAK01} proposed to change the initial
conditions by changing the wavelength of the light source. Umeno
et al. \cite{UAK01} performed experiments for the cases $m=2$ and
$m=3$, and $N=2$. They measured the values $X_1$ and $X_2$ several
times changing the wavelength in the range from $1560.00$ nm to
$1560.5$ nm. The figures they obtained are approximately
equivalent to the first--return maps of the logistic and the cubic
maps (See Fig. 1.)

We should stress here that any chaotic map of type
$X_{n+1}=f(X_n)$ is predictable in the short term because
$X_{n+1}$ is always defined as a function of the previous value.
For instance, for the logistic map, whenever $X_n\approx 0.1$,
$X_{n+1}\approx 0.36$.

In the present Letter, we will show that functions of type
\begin{equation}
X_n = P[\theta T Z^n],
\label{EQ4}
\end{equation}
where $P[t]$ is a periodic function, $T$ is the period of $P[t]$,
$\theta$ is a real parameter and $Z$ is a noninteger number, can
generate random dynamics in the sense that $X_{n+1}$ is not
determined by any string $X_0, X_1,\ldots,X_n$ of previous values.

Moreover we will show that the sequence of values $X_n$ is such
that past values cannot be used to predict future values and
future values cannot be used to ``predict'' past values.
Furthermore, for irrational $Z$, there are no correlations at all
between the values of the sequences.

Let us define the family of sequences
\begin{equation}
X_{n}^{k,m,s}:=P\left[T\left(\theta_0 + q^m
k\right)\left(\frac{q}{p}\right)^s\left(\frac{p}{q}\right)^n\right],
\label{EQ5}
\end{equation}
where $k$, $m$ and $s$ are integer. The parameters $k$
distinguishes the different sequences. For all sequences
parametrized by $k$, the strings of $m+1$ values
$X_s,X_{s+1},X_{s+2},\ldots,X_{s+m}$ are the same. This is so
because $X_{n}^{k,m,s}=P\left[T\theta_0(q/p)^s(p/q)^n\right]$, for
all $s\leq n \leq m+s$. So we can have an infinite number of
sequences that share the same string of $m+1$ values.
Nevertheless, the next value $X_{s+m+1}^{k,m,s}=
P\left[T\theta_0(p/q)^{m+1}+Tkp^{m+1}/q\right]$ is uncertain. In
general, $X_{s+m+1}^{k,m,s}$ can take $q$ different values. In
addition, the value $X_{s-1}$, ($X_{s-1}=P\left[ T\theta_0(q/p) +
T k q^{m+1}/p \right]$), is also undetermined from the values of
the string $X_{s},X_{s+1},X_{s+2},\ldots,X_{s+m}$. There can be
$p$ different possible values for $X_{s-1}$. Thus, for any string
$X_{s},X_{s+1},X_{s+2},\ldots,X_{s+m}$, the future and the past
are both uncertain. In the case of a generic irrational $Z$, there
are infinite possibilities for the future and the past.

We should remark that the functions (\ref{EQ4}), with a noninteger $Z=p/q$,
are not in the class $X_{n+1}=f(X_n)$ nor $X_{n+1}=f(X_n,\ldots,X_{n-m+1})$, however they can be expressed as
random maps of type $X_{n+1}=f(X_n, I_n)$, where $I_n$ is a time--dependent random variable
(See Refs.\cite{GRG01,GTA03}). That is, the fact that $Z$ is not integer leads to a sort of time dependence.

Some properties of function $X_n = \sin^2[\theta \pi Z^n]$, which
is a particular case of (\ref{EQ4}), have been already studied in
previous papers (See e.g. \cite{GRG01,GTA03,GC97,GMLT00,GR01}.)

Here we will show that there are no statistical correlations
between $X_n$ and $X_m$ (where $n\neq m$). We will investigate the
functions $U_n=\cos(\theta \pi Z^n)$, which possess zero mean.
Note that $X_n = 1 - U_n^2$. The values of $U_n$ are found in the
interval $-1\leq U_n \leq 1$. Let us define the $r$--order
correlations\cite{Be91,HB01}:
\begin{equation}
E(U_{n1}U_{n2}\cdots
U_{nr}):=\int_{-1}^{1}\textrm{d}U_0\left[\rho(U_0)U_{n1}U_{n2}\cdots
U_{nr}\right].
\end{equation}
The invariant density $\rho(U)$ is given by
$\rho(U)=1/(\pi\sqrt{1-U^2})$ and $U_0= \cos(\pi \theta)$. We have
the following formula for the correlation functions
%
%
$E(U_{n_1}U_{n_2}\cdots U_{n_r})=$\\  $\int_0^1
\textrm{d}\rho\left[\cos(\theta\pi Z^{n_1}) \cos(\theta\pi
Z^{n_2}) \cdots \cos(\theta\pi Z^{n_r})\right]$.
%
%
Considering that $\cos \theta =
\frac{1}{2}\left(e^{i\theta}+e^{-i\theta}\right)$, we obtain
\begin{equation}
E(U_{n_1}U_{n_2}\cdots
U_{n_r})=2^{-r}\sum_{\sigma}\delta(\sigma_1 Z^{n_1}+\sigma_2
Z^{n_2}+ \cdots + \sigma_r Z^{n_r},0),
\end{equation}
where $\sum_{\sigma}$ is
the summation over all possible configurations $(
\sigma_1,\sigma_2,\ldots,\sigma_r )$, with $\sigma = \pm 1$, and
$\delta (n,m)=1$, if $n=m$ or $\delta (n,m)=0$, if $n\neq m$. We
will have non--zero correlations only for the sets
$(n_1,n_2,\ldots,n_r)$ that satisfy the equation
\begin{equation}
\sum_{i=1}^{r}\sigma_iZ^{n_i}=0,
\end{equation}
where $\sigma_i = \pm 1$.

It is easy to see that, for $Z>1$,
\begin{equation}
E(U_nU_{m})=0
\end{equation}
if $n\neq m$.
As a particular case $E(U_nU_{n+1})=0$. We also wish to show that
the correlation functions $E(U_n^{i}U_{n+1}^{j})$ are zero when
$i$ is even and $j$ is odd (or vice versa) i.e.
\begin{equation}
E(U_n^{i}U_{n+1}^{2j + 1})=0.
\end{equation}
So let us calculate
$E(U_n^{i}U_{n+1}^{2j + 1})$: $E(U_n^{i}U_{n+1}^{2j +
1})=\frac{1}{2^r}\sum_{\sigma}\delta (\sigma_1Z^{n}+\cdots +
\sigma_iZ^{n}+\sigma_{i+1}Z^{n+1}+\cdots +\sigma_{i+2j
+1}Z^{n+1},0)$. Note that $E(U_n^{i}U_{n+1}^{2j + 1})$ is not zero
only if $(\sigma_1+\sigma_2+\cdots +
\sigma_i)Z^{n}+(\sigma_{i+1}+\sigma_{i+2}+\cdots
+\sigma_{i+2j+1})Z^{n+1}=0$. For irrational $Z$, this equation has
not solutions.

Now we wish to consider all the possible correlations
$E(U_{n_1}U_{n_2}\cdots U_{n_r})$. We should note that $E$ can be
non--zero in some ``trivial'' cases (for instance $E(U_n^{2j})$,
$E(U_n^{2j}U_{n+1}^{2j})$) which are related to moments
$E(U_n^{2j})$. This does not affect randomness \cite{Be91,HB01}.
In the language of equations
$\sum_{i=1}^{\sigma}\sigma_iZ^{n_i}=0$, this can happen only due
to trivial cases as the following $Z^n - Z^{n}-Z^{n+1}+Z^{n+1}=0$.

We will show that all the ``nontrivial'' correlations are zero for
our functions. Suppose $n_r=2j+1$. Then $E(U_{n_1}U_{n_2}\cdots
U_{n_r})$ is not zero only if there are solutions for the
equations $\sum_{i=1}^r \sigma_i Z^{n_i}=0$. But these equations
can be written in the form
\begin{equation}
N_0 + N_1 Z + \cdots + N_{2j + 1}Z^{2j
+1}=0,
\end{equation}
where $N_i$ are integer, and $N_{2j+1}\neq 0$. For
transcendent irrational $Z$ this equation is never satisfied. Thus
the sequences generated by function $X_n = \sin^2[\theta\pi Z^n]$
with a transcendent $Z$ are completely uncorrelated.

Different aspects of the predictability problem are used in references \cite{Block1,CFOKV00,BCFV02,KO00}
as a way to characterize  complexity and to find distinction between noise and chaos
in experimental time series.

In Refs. \cite{Block1,CFOKV00,BCFV02,KO00} several quantities are introduced in order to determine the true
character of the time series. All these methods have in common that one has to choose
certain length scale $\epsilon$ and a particular embeding dimension $m$. The mentioned
quantities discussed in these articles dsiplay different behaviors as the
resolution is varied. According to these different behaviors one can distinguish chaotic
and stochastic dynamics.

Using the results of Refs.\cite{GRG01,GTA03,GC97,GMLT00,GR01} and the present paper, it is
possible to prove that functions (\ref{EQ4})
can represent different kinds of dynamics: chaotic time series (with integer $Z$),
random maps or unpredictable sequences (with noninteger $Z$), and completely uncorrelated sequences of independent values (with generic irrational $Z$.)

Our functions can be investigated analytically and their complexity can be calculated exactly using
theoretical considerations \cite{GRG01,GTA03}. So these functions can be used as very suitable models in order
to check the predictions of Refs.\cite{Block1,CFOKV00,BCFV02,KO00}.

Moreover, we can produce long sequences of values using our models and then, we can study them as
experimental time series in the framework of the methods presented in Refs.\cite{Block1,CFOKV00,BCFV02,KO00}.
In fact, we have investigated the asymptotic behavior of the quantities discussed
in Ref.\cite{Block1,CFOKV00,BCFV02,KO00},
and our results coincide with those obtained in the mentioned papers.
Additionaly, we have checked the formula $K=\lambda \theta (\lambda) + h$ for the
complexity of random maps of type $X_{n+1}=f(X_n,I_n)$, where $K$ is the complexity of the system,
$\lambda$ is the Lyapunov exponent of the map and $h$ is the complexity of $I_n$, and
$\theta(\lambda)$ is the Heaviside step function \cite{BCFV02}.
Details of the applications of our results in the problem of distinguishing a chaotic system
from one with intrinsic randomness will be presented in a more extended paper.

The statistical properties of pseudo--random number generators are discussed in Refs.\cite{Block2}. The authors of
these papers have noticed that almost all pseudo--random number generators calculate a new pseudo--random number $X_{n+1}$ using a recursive formula that depends on the preceeding values $X_{n+1}=f(X_n, X_{n-1}, \ldots,X_{n-r+1})$. They have found that the failure of these generators in different simulations can be attributed to the low entropy of the production rule $f()$ conditioned on the statistics of the input values  $X_n, X_{n-1}, \ldots,X_{n-r+1}$. Besides, all these generators have very strong correlations even at the macrostate level used in the simulations \cite{Block2}.

We agree with these researchers that this approach, based on the properties of the generator rule, is more profound than the empirical tests.

In this same spirit, we should say that the rule (\ref{EQ4}) produces a dynamics where
the future values are not determined by the past values. In fact, they can be completely unconrrelated.

Now suppose that we have the same experimental setup of Ref.
\cite{UAK01}, which is represented in Fig. 2 schematically, but
the equation for the path length differences $\Delta L(n)$ will
satisfy a relationship very similar to Eq. (\ref{EQ1}), that is
$\Delta L(n+1)=Z\Delta L(n)$, with the change that $Z$ is not an
integer. In this case the sequence $X_n$ of measured light
intensities will be unpredictable.

\begin{figure}
\onefigure[width=8cm,height=4cm]{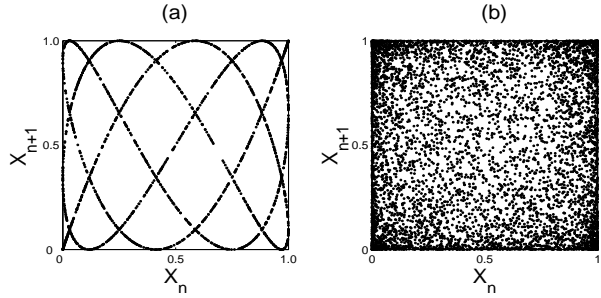}
\caption{First--return maps produced by function
$X_n=\sin^2[\theta\pi Z^n]$ with $Z=1.8$ (a) and $Z=\pi$ (b).}
\label{f.2}
\end{figure}

Fig. \ref{f.2} shows different examples of the
dynamics that can be produced by function $X_n=\sin^2[\theta\pi
Z^n]$ with different noninteger  $Z$.

Using the properties of function (\ref{EQ4}) and further
investigation we can obtain the following results. Function
$X_n=P[\phi(n)]$, where $P[t]$ is a periodic function and
$\phi(n)$ is a non--periodic oscillating function with
intermittent intervals of truncated exponential behavior, would
produce also unpredictable dynamics.

Furthermore, we can construct functions of type $X_n=h[\phi(n)]$,
again with very complex behavior, where $h(t)$ is a
non--invertible function and $\phi(n)$ is, as before, a
non--periodic oscillating function with intermittent intervals of
truncated exponential behavior. Some chaotic systems can produce
the kind of behavior needed for $\phi(n)$. This physical system can be constructed, for
example, with circuits: a chaotic circuit and a circuit with a
non--invertible $I$--$V$ characteristic \cite{CDK87}.
An experiment  with this scheme is reported in \cite{GRSGG02}.

\begin{figure}
\onefigure[width=8cm,height=3cm]{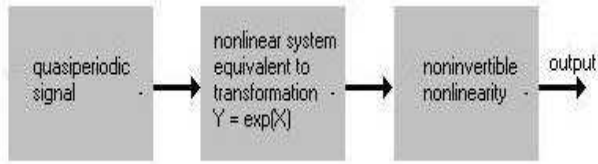}
\caption{Experimental setup to produce unpredictable dynamics
using a quasiperiodic signal.}
\label{s.3}
\end{figure}

However the most interesting fact is that we can construct
function $\phi(n)$ without using previously produced chaotic
signals. In this case we plan to use as input to the non--linear
system only regular signals.

A consequence of our theory is that a time--series constructed
using three periodic signals can be transformed into an
unpredictable dynamics. The theoretical result is that the
following function can be unpredictable:
\begin{equation}
X_n = P[A\exp [Q(n)]],
\label{EQ6}
\end{equation}
where $P[y]$ is a periodic function (in some cases it can be just
a non--invertible function,) and $Q(n)$ is a quasiperiodic
function represented by the sum of several periodic functions.

As an illustrating example let us study the following function
\begin{equation}
X_n = \sin[\phi(n)],
\label{EQ7}
\end{equation}
where $\phi(n) = A\exp[Q(n)]$, $Q(n)=P_1(n)+P_2(n)+P_3(n)$,
$P_i(n) = a(n-kT_i)$, when $kT_i\leq n \leq (k+1)T_i$. Here
$T_2/T_1$, $T_3/T_2$ and $T_3/T_1$ are irrational numbers.

Note that the functions $P_i(n)$ are piece--wise linear. Function
$Q(n)$ is also piece--wise linear, but it is not periodic. On the
other hand, function $\phi(n)$ will behave as a non--periodic
oscillating function with intermittent intervals of finite
exponential behavior. At these intervals, function $X_n$ behaves
as function $X_n=\sin^2[\theta \pi Z^n]$.

Similar properties can be found in function (\ref{EQ7}) if $Q(n) =
a_1\sin(\omega_1 n)+ a_2\sin(\omega_2 n)+a_3\sin(\omega_3 n)$. In
fact, functions $\sin(\omega_i n)$ behaves approximately as
increasing linear functions whenever $\omega_n\approx 2\pi k$
where $k$ is an integer. We have performed real experiments using
the setup represented in Fig. \ref{s.3}.

\begin{figure}
\onefigure[width=4cm,height=4cm]{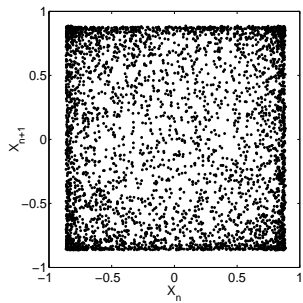}
\caption{Dynamics produced by the experiment.}
\label{f.4}
\end{figure}

In our experiments, a quasiperiodic time--series was used as input
to an electronic circuit that simulates an exponential function
\cite{S76}. The output of the exponential system is taken as the
input to a non--linear system that simulates the sine--function
\cite{Ma78}. Fig. \ref{f.4} shows an example of the dynamics
produced by the experiment. Details of the experiment will be
presented elsewhere in a more extended paper.

In conclusion, we have shown that functions of type $X_n=P[\theta
T Z^n]$, where $P[t]$ is a $T$--periodic function, $\theta$ and
$Z$ are real numbers, can generate random dynamics in the sense
that any string of values $X_{s},X_{s+1},\ldots,X_{s+m}$ is
deterministically independent of past and future values.
Furthermore, there are no correlations whatsoever between the
values of the sequence.

The experimental setup schematically represented in Fig. 2 (See
Ref. \cite{UAK01}), where the path--length differences in the
Mach--Zehnder interferometers satisfy the equation $\Delta
L(n+1)=Z\Delta L(n)$ (with noninteger $Z$), can be used to produce
unpredictable dynamics.

We have performed real experiments with systems that are
equivalent to the scheme represented in Fig. \ref{s.3}. These
experiments corroborate our prediction that, using just static
non--linear systems, a quasiperiodic signal can be transformed
into a random signal.


\end{document}